\documentclass[twocolumn,pre,superscriptaddress,epfs,aps,showpacs]{revtex4}
\usepackage{amsmath}
\usepackage{amssymb}
\usepackage{graphicx}
\usepackage{color}
\usepackage[dvipsnames]{xcolor}
\usepackage{epsfig}
\usepackage{braket}	
\usepackage{latexsym}
\usepackage{tikz,siunitx}
\usepackage{scalerel}
\usepackage{etoolbox}
\usetikzlibrary{shapes.geometric,shapes.symbols}
\definecolor{myyellow}{RGB}{247, 199, 89}
\definecolor{myred}{RGB}{230, 99, 79}
\definecolor{myblue}{RGB}{154, 202, 202}
\definecolor{mygreen}{RGB}{185, 181, 120}
\definecolor{mypink}{RGB}{240, 226, 231}
\definecolor{mydarkblue}{RGB}{95, 104, 150}
\definecolor{myautumm}{RGB}{198, 178, 153}
\definecolor{mycolornude}{RGB}{230, 213, 193}
\definecolor{mywhitepeach}{RGB}{87, 10, 63}
\definecolor{mymaroon}{RGB}{11, 72, 107}

\newcommand{\mytriangle}[1]{\tikz{\node[draw=black,fill=#1,isosceles
triangle,isosceles triangle stretches,shape border rotate=90,minimum
width=0.3cm,minimum height=0.3cm,inner sep=0pt] at (0,0) {};}}

\newcommand{\mytrgle}[1]{\tikz{\node[draw=black,fill=#1,isosceles
triangle,isosceles triangle stretches,shape border rotate=90,minimum
width=0.2cm,minimum height=0.2cm,inner sep=0pt] at (0,0) {};}}
\newcommand{\mysquare}[1]{\tikz{\node[draw=black,fill=#1,rectangle,minimum
width=0.3cm,minimum height=0.3cm,inner sep=0pt] at (0,0) {};}}

\newcommand{\mycircle}[1]{\tikz{\node[draw=black,fill=#1,circle,minimum
width=0.3cm,minimum height=0.3cm,inner sep=0pt] at (0,0) {};}}

\newcommand{\mycirc}[1]{\tikz{\node[draw=black,fill=#1,circle,minimum
width=0.2cm,minimum height=0.2cm,inner sep=0pt] at (0,0) {};}}

\begin{document}

\title{Emergence of Synchronization-Induced Patterns in Two-dimensional Magnetic Rod Systems under Rotating Magnetic Fields }

\author{Jorge L. C. Domingos}
\affiliation{Departamento de F\'{i}sica, Universidade Federal do Cear\'{a}, Fortaleza, Cear\'{a}, Brazil}
\author{F. Q. Potiguar}
\affiliation{Universidade Federal do Par\'a, Faculdade de F\'\i sica, ICEN, Bel\'em, Par\'a, Brazil}
\author{C. L. N. Oliveira}
\affiliation{Departamento de F\'{i}sica, Universidade Federal do Cear\'{a}, Fortaleza, Cear\'{a}, Brazil}
\author{W. P. Ferreira}
\affiliation{Departamento de F\'{i}sica, Universidade Federal do Cear\'{a}, Fortaleza, Cear\'{a}, Brazil}

\date{\today}

\begin{abstract}
We investigate the dynamics of two-dimensional assemblies of rod-shaped magnetic colloids under the influence of an external rotating magnetic field. Using Molecular Dynamics, we simulate the formation of patterns that emerge based on the synchronization degree between the magnetic rods and the rotating field. We then explore the structural and dynamic characteristics of the resulting steady states, examining their evolution as a function of changes in the rods' aspect ratio, the strength of the external magnetic field, and its rotation frequency. Three distinct synchronization regimes of the rods with the magnetic field are clearly observed. A detailed set of phase diagrams illustrates the complex relationship between the magnitude of the external magnetic field and its rotation frequency and how these parameters govern the formation of unique self-organized structures.
\end{abstract}


\maketitle

\section{Introduction}
The self-organization of magnetic particles (MP) guided by rotating fields has been realized as a powerful tool to generate functional materials \citep{SYAN,YAN,usulis}. Microstructures of MP in the presence of external fields form a rich non-equilibrium phase diagram composed of chains, clusters, crystals, vortices, and many other configurations \citep{JorgeSM,Dobnikar,JAMES,Belovs},  drawing enormous attention in recent years due to their relevant applications, such as measuring biomolecular interactions \cite{Lchen,Gupta}, developing microfluidic devices \citep{Bleil,Tasci,shanko}, tuning magnetorheological suspensions \cite{Vicente,terkel}, and colloidal assemblies \citep{JYAN,Salazar,JyanSM,Swanet}. Time-dependent fields are an essential tool for controlling self-organization processes; thus an increasing interest has been devoted to the collective behavior of magnetic particles subject to external magnetic fields. Previous experiments and computer simulations reveal that, for strong enough rotating fields, spatial symmetry can be broken by the formation of layers in the field plane \citep{Martin,Muller,SJager,Elsner,Martin2}.

Many magnetic field-driven pattern formation phenomena can be explained from an equilibrium perspective if the particles possess induced dipoles. However, less is known about the corresponding behavior of particles with permanent dipole moments, such as ferromagnetic particles, in response to rotating fields. In this case, their individual orientations may differ from that of the rotating field due to hydrodynamic and thermal effects. Therefore, the essential prerequisite for full synchronization with the field, thus building a time-averaged potential, may not be fulfilled \cite{klapp2}. Here, the particles rigidly rotate because of the magnetic torque on the dipoles, the so-called spinners. The collective behavior of spinners is governed by magnetic and hydrodynamic interactions \cite{Junot}. By tuning these interactions, we can probe the phase diagram of this system and study the emergent collective dynamics and pattern formation. 

Further interesting phenomena occur when the spinners are exposed to rotating fields in two-dimensional geometry, since the time-averaged dipolar interaction is purely attractive and long-ranged \citep{klapp2,Junot,Jaeger}. As a result, spatially non-symmetric ensembles of dynamical clusters, namely swarms, are observed above a critical field intensity and below a critical frequency of the magnetic field behaving as liquid droplets. Recent studies have discussed biomedical applications where such swarms could be functional in performing adaptive pattern control in complex environments, providing targeted delivery with a high access rate \cite{Jiangfan}, and operating as selective actuation in Magnetic Particle Imaging (MPI) technique \cite{Bente}. Soni et al. \cite{Vishal} used magnetic-responsive cubic spinners to engineer chiral fluids to study odd viscosity. 
Such field-induced aggregations tend to be formed with different sizes, patterns, and angular velocities \cite{Petrichenko}. The complete comprehensive understanding of pattern control of such aggregates in the presence of rotating fields remains challenging, and a profound investigation is lacking.

The synchronization-driven pattern formation primarily centers on the external field parameters, i.e., the field intensity and rotation frequency, as in the particle interaction features \cite{SJager,JorgeSM}. However, the effect of the particle geometry anisotropy on the resultant phase diagrams has yet to be explored. Here, we employ a peapod-like model to simulate rigid magnetic rods under an in-plane rotating magnetic field. Such a model, considered experimentally \cite{birringer2008magnetic} and numerically \citep{JorgeSM,alvarez,jdomingos2,jorgedomingos}, allow us to explore the interplay between the aspect ratio of the ferromagnetic rods and the external magnetic field on the resultant synchronization-driven pattern formation. The extent of such synchronization leads to emergence of different patterns, a result of the interplay between the rod-field and rod-rod interactions, which can be tuned by adjusting the aspect ratio of the rods, field intensity, and rotation frequency. Friction between rods and solvent also plays an important role in the pattern formation. Moreover, the external magnetic field may induce an effective paramagnetic or diamagnetic order of the system depending on the rotation frequency of the external field. This complexity underscores the extent of our investigation and the challenges we aim to overcome.

\section{Model}\label{model}
\begin{figure}[h!]
\centering
 \includegraphics[height=6cm]{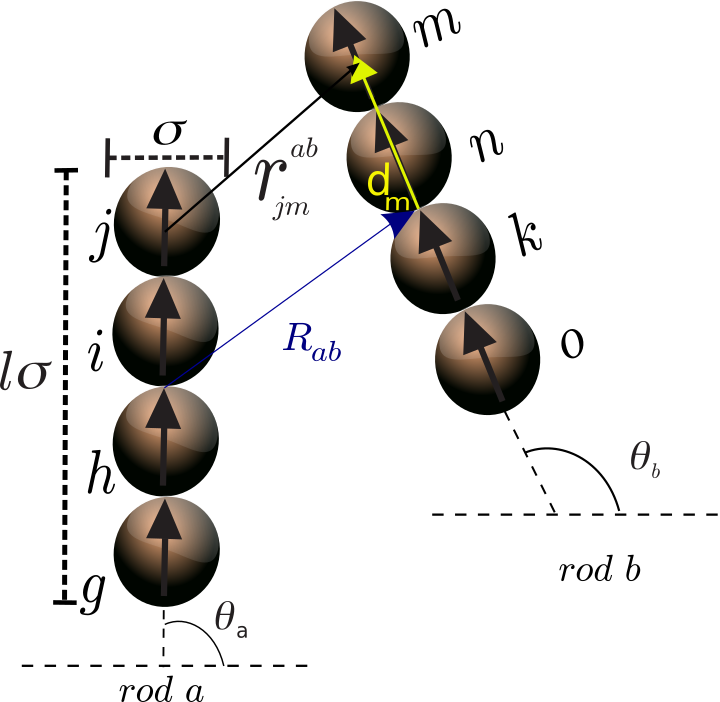}
  \caption{ Schematic illustration of the interaction between two magnetic rods with an indication of the important parameters of the pair interaction potential.}
  \label{Fig1}
\end{figure}

We use molecular dynamics simulations \cite{PhysRevResearch.2.033222, 10.1371/journal.pone.0299296} to study a two-dimensional ($2D$) system consisting of $N$ rigid peapod-like rods of aspect ratio $l$. We simulate the rod's magnetic nature by attaching a point-like dipole of permanent magnetic moment $\mu$ at the center of each bead (see Fig. \ref{Fig1}).

The orientation of the dipoles is along the axial direction of the rod, as illustrated in Fig. \ref{Fig1}. To model the interaction between dipolar particles, we use a dipolar soft sphere (DSS) potential \cite{stevens}, consisting of the repulsive part of the Lennard-Jones (LJ) potential, $u^{rep}$, and a point-like dipole-dipole interaction, $u^{D}$. The total interaction energy between rods $a$ and $b$ is the sum of the pair interaction terms between their respective dipolar spheres (DS)

\begin{eqnarray}
U_{a,b}(\mathbf{R}_{a,b},\theta_a,\theta_b) = \sum_{j \neq m} u_{j,m}\textrm{,} \\
u_{j,m} =  u^{rep}(\mathbf{r}_{jm}^{a,b}) + u^{D}(\mathbf{r}_{jm}^{a,b},\boldsymbol{\mu}_j^{a},\boldsymbol{\mu}_m^{b})\textrm{,}  
\end{eqnarray}
where

\begin{eqnarray}
u^{rep} &=& 4\epsilon \left(\frac{\sigma}{r_{jm}}\right)^{12}\textrm{,}  \\
u^{D} &=& \frac{\boldsymbol{\mu}_{j}\cdot\boldsymbol{\mu}_{m}}{r_{jm}^3}-
\frac{3(\boldsymbol{\mu}_{j}\cdot\mathbf{r}_{jm})(\boldsymbol{\mu}_{m}\cdot\mathbf{r}_{jm})}{r_{jm}^5}\label{eqdip}\textrm{,}
\end{eqnarray}

\noindent with $\sigma$ the diameter of each bead, and $\epsilon$ the LJ soft-repulsion constant. $\mathbf{R}_{a,b} = \mathbf{R}_b - \mathbf{R}_a$ is the vector that locates the center of rod $b$ relative to rod $a$'s center. The orientations of rods $a$ and $b$ are given by the angles $\theta_a$ and $\theta_b$, respectively. The vector $\textbf{r}_{jm}^{a,b}$ locates the center of bead $m$ of rod $b$ concerning the center of a given bead $j$ of rod $a$ (see Fig. \ref{Fig1}). The force on bead $m$ due to bead $j$ is given by

\begin{equation}
\mathbf{f}_{jm} = - \boldsymbol{\nabla} u_{jm}\textrm{.}
\end{equation}

The torque on bead $m$ is \cite{jorgedomingos}
\begin{eqnarray}\label{torque1}
\boldsymbol{\tau}_{m} = \boldsymbol{\mu}_m \times \sum_{m\neq j}\lbrace\mathbf{B}_{jm}+\mathbf{B}(t)\rbrace + \mathbf{d}_m \times \sum_{m\neq j}\mathbf{f}_{jm}\textrm{,}
\end{eqnarray} 
\noindent where $\textbf{d}_m$ is the vector connecting the center of bead $m$ (rod $b$) with the center of rod $b$, as illustrated in Fig. \ref{Fig1}, and $\textbf{B}_{jm}$ is the magnetic field generated by the dipole moment $\mu_j$ at the position of the dipole $\mu_m$, and $B(t)$ is the external magnetic field. They are given by
\begin{eqnarray}
\mathbf{B}_{jm} = \frac{3(\boldsymbol{\mu}_{m}\cdot\mathbf{r}_{jm})\mathbf{r}_{jm}}{r_{jm}^5}-\frac{\boldsymbol{\mu}_{m}}{r_{jm}^3}\textrm{,}\\
\mathbf{B}(t) = B_0[\cos(\omega_0 t)\mathbf{\hat{x}}+ \sin(\omega_0 t)\mathbf{\hat{y}}]\textrm{,}
\end{eqnarray}
\noindent where $B(t)$ is the external magnetic field, with intensity $B_0$ and rotation frequency $\omega_0$ (also mentioned as driving frequency in the text). The external magnetic field rotates in the same plane as the magnetic rods.

 The orientation of the rods is given by the unitary vector $\mathbf{\hat{s}}$ given by, for any $m$ on a given rod, $\mathbf{\hat{s}} = \textbf{d}_m/\mid{\textbf{d}_m}\mid$. The translational and rotational Langevin equations of motion of a rod with mass $M$ and moment of inertia $I$, are given by
\begin{eqnarray}
M\frac{d\mathbf{v}}{dt} = \mathbf{F} - \boldsymbol{\Gamma_T}\cdot\mathbf{v}+ \boldsymbol{\xi }^T(t)	\textrm{,}
\\
I\frac{d\mathbf{\boldsymbol\omega}}{dt} = \mathbf{N} - \zeta_r \boldsymbol{\omega} + \boldsymbol{\xi }^R(t)\textrm{,}\label{eqrot}
\end{eqnarray}  

\noindent where $\mathbf{v} = d\mathbf{R}/dt$, $\boldsymbol{\omega }$ is the angular velocity of the rod,  $\mathbf{F}$ and $\mathbf{N}$ are the total force and total torque, respectively. At the same time, while $\boldsymbol{\Gamma_T}$ and $\zeta_r$ are the translational friction tensor and rotational friction parameters, respectively. For rod-like particles, the translational tensor is composed of the parallel ($\zeta_{\parallel}$) and perpendicular ($\zeta_{\perp}$) components relative to the rod axis, which are given by
\begin{equation}
\zeta_{\parallel}= \frac{2\pi\eta_0l\sigma}{ln(l) + \delta_{\parallel}}\textrm{,}\quad\zeta_{\perp}= \frac{4\pi\eta_0l\sigma}{ln(l) + \delta_{\perp}}\textrm{,}\label{eqfri1}
\end{equation} 
\noindent and for rotation 
\begin{equation}
\quad\zeta_r =  \frac{\pi \eta_0 ({l\sigma})^3}{3\,ln(l)+\delta_r}\textrm{,}\label{eqfri2}
\end{equation}
\begin{figure*}[t]
\centering
   \hspace{-10.pt}\includegraphics[height=9.cm]{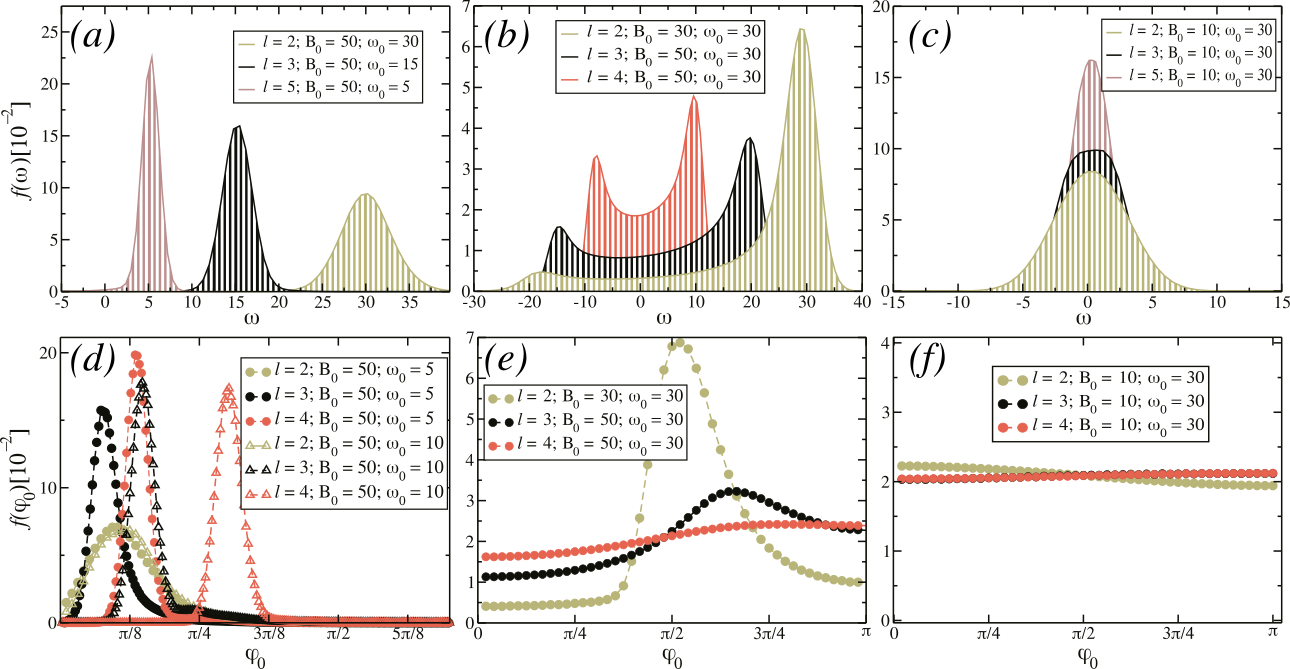}
  \caption{Distribution of frequencies and phase differences for: High Synchronization regime (a) and (d), respectively; Intermediate Synchronization regime (b) and (e), respectively; and Low Synchronization regime (c) and (f), respectively.} 
  \label{freqdist}
\end{figure*}
\noindent where $\eta_0$ is the solvent viscosity, $\delta_{\parallel}$, $\delta_{\perp}$, and $\delta_r$ are correction factors for small rods extracted from Refs. \cite{tirado1,tirado2}. As a result, the total translational diffusion coefficient is $D_T = \frac{1}{3}(D_{\parallel}+2D_{\perp})$ for $D_{\perp} = \frac{1}{2}D_{\parallel}$ \cite{langrods}.
$\boldsymbol{\xi }^T$ and $ \boldsymbol{\xi }^R$ are the Gaussian random force and torque, respectively, which obey the following white noise conditions for a given rod $b$ at time $t$: $\braket{\boldsymbol{\xi }^\alpha_b(t)} = 0$, $\braket{\boldsymbol{\xi }^\alpha_b(t)\cdot\boldsymbol{\xi }^\alpha_{b'}(t')}=2\Gamma_\alpha k_BT\delta_{bb'}\delta(t-t')$, $\alpha = T,R$, where $\delta_{bb'}$ and $\delta(t-t')$ are the Kronecker and Dirac delta functions, respectively.   

We define the reduced unit of time as $t^*=t/\sqrt{\epsilon^{-1}M\sigma^2}$, where $M$ is the mass of the rod. Therefore, the rotation frequency of the external magnetic field is $\omega_0^*=\omega_0\sqrt{\epsilon^{-1}M\sigma^2}$. The dimensionless energy, dipole moment, distances, and magnetic field are given as $U^* = U/\epsilon$, $\mu^{*} = \mu\sqrt{10^{-7}\epsilon \sigma^3}$, $r^* = r/\sigma$, $B_0^* = B_0/\sqrt{10^{-7}\epsilon/\sigma^3}$, respectively. We set  $k_BT/\epsilon = 1$, such that $\epsilon/k_B$ is the temperature unit and $k_B$ is the Boltzmann constant. Periodic boundary conditions are taken in both spatial directions. No special long-range summation techniques are needed since the dipolar interaction falls off as $(r^{-3})$, and we take the simulation box large enough so that the interaction between two particles separated $L/2$ from each other is negligible \cite{allen1989computer}. We define the packing fraction as $\eta = N_{beads}\pi(\sigma/2)^2/L^2$, where $N_{beads} = 3000$ is the total number of dipolar beads of the system, and $L^2$ is the simulation box area. Since $N_{beads} = lN$, we can rewrite the packing fraction as $\eta = \rho^*l \pi/4$, where $\rho^*$ is the dimensionless density $\rho^* = \rho\sigma^2$, and $\rho = N/L^2$, in all simulations, we set $\eta = 0.1$. 

The reduced time step is in the range $\delta t^* =  10^{-4} - 10^{-3}$.
The quantities of interest are averaged over $4 \times 10^6$ time steps. 
We study the pattern formation as a function of the aspect ratio $l$, the intensity of the external magnetic field $B_0$, and its rotation frequency $\omega_0$. For the former, we use $l = 2, 3, 4$ and $5$. The total dipole moment of each rod is $\displaystyle{\sum^l\mu^* = 13.2}$, regardless of its aspect ratio. Thus, the dipole moment of a single bead is in the range of $2.64 \leq \mu^* \leq 6.6$. For instance, for $l = 3$, we have $\mu^*=4.4$, which was considered in a previous study \cite{JorgeSM}, and it was estimated from experiments at room temperature ($T \approx 293 K$) using iron nanoparticles \cite{birringer2008magnetic} with saturation magnetization $M_s(Fe) = 1700$ $kA/m$ and radius $r \approx 5$ $nm$.
For the external fields, we use $B_0^*$ values within the range $10\leq B_0^* \leq 50$, related to the magnetic field at room temperature at $33$ $mT$ $\leq B_0 \leq 165$ $mT$. Experimental values for the magnetic fields are of the order of $0.1$ $T$ \cite{alvarez}, but ferrofluids are already susceptible to $B_0 < 10$ $mT$ \cite{birringer2008magnetic}. We are omitting the superscript $^*$ hereafter in all dimensionless parameters for simplification. In the following section, we present our numerical results whose analysis is based on quantities that will be presented in the next section.

\section{Results and Discussion}\label{results}
In this section, we discuss the system's synchronization regimes and pattern formation as a function of the field intensity $B_0$ and driving frequency $\omega_0$ for rods with different aspect ratios $l$. We characterize the rod's aggregation state through the pair correlation function. We also define a proper correlation function to study the rods' synchronization and time-spatial ordering.

\begin{figure}[]
 \hspace{-40.pt} \includegraphics[height=7.5cm]{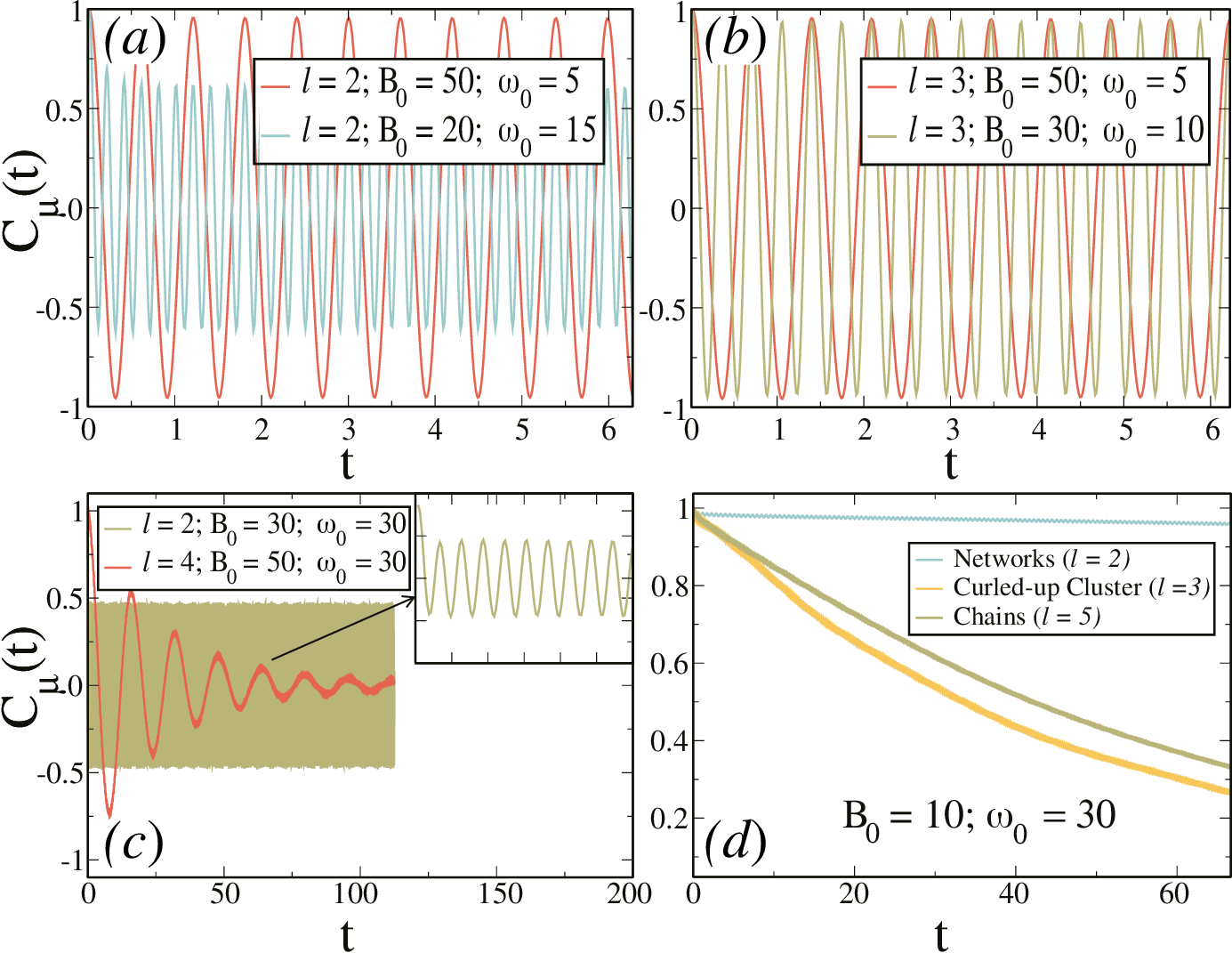}
 \hspace{0.pt} \caption{Examples of dipole autocorrelation functions for the different regimes of synchronization: High Synchronization regime for (a) $l = 2$, and (b) $l = 3$; (c) Intermediate Synchronization regime; and (d) Low Synchronization regime for the observed clustered phases: Networks ($l = 2$); Curled-up Cluster ($l = 3$) and Chains ($l = 5$) (a comprehensive analysis of these findings will be discussed in section $3.2.3$). The inset in (c) is a zoom-in illustration of the corresponding curve.}
  \label{fgr:DAF}
\end{figure}
\subsection{Determination of the synchronization regime }
The synchronization of the rods with the external magnetic field, hereafter magnetic field or simply field, plays an important role in determining the different pattern formation. The synchronization regimes are related to the rotation of rods in response to the magnetic field. This behavior is quantitatively based on the analysis of the normalized distributions of angular velocities of the rods ($f(\omega)$), and phase difference ($f(\varphi_0)$) between each rod and the magnetic field (Fig. \ref{freqdist}). More precisely
\begin{equation}\label{dist}
f(\alpha) = \frac{1}{N}\Braket{\sum_{i}^N\delta(\alpha -\alpha_i)},
\end{equation}
where $\delta(\alpha -\alpha_i)$ is the Dirac delta function, and $\alpha = \omega, \varphi_0$. In addition, the analysis of the rotation dynamics of the dipoles by using the single dipole autocorrelation function (DAF) also provides valuable insights about the regime of synchronization, and it is given by
\begin{equation}\label{eqcorr}
C_{\mu}(t) = \frac{1}{N}\Braket{\sum_{i= 1}^{N}\hat{\mu}_{i}(t)\cdot\hat{\mu}_{i}(0)}\textrm{,}
\end{equation}
where $N$ is the total number of rods, $\hat{\mu}$ is the unitary vector pointing in the direction of the magnetic moment of the $i$-th rod at time t, and $\Braket{ \quad}$ represents the average over time (the same notation is used throughout the manuscript). Depending on the synchronization regime, the DAF may present a free harmonic oscillation, a damped oscillation, or even a monotonic decay as a function of time. 

In the High Synchronization (HS) regime, the angular velocities $\omega$ of most particles are equivalent to the driving frequency of the magnetic field $\omega_0$. In this case, the distribution $f(\omega)$ presents a single peak centered at $\omega \approx \omega_0$ [Fig \ref{freqdist}(a)]. Also, the DAF results in \cite{JorgeSM}
\begin{equation}\label{eq:amplitudecorr}
C_{\mu}(t) = \frac{n_s}{N}\cos(\omega_0 t)\textrm{,}
\end{equation}
\noindent where $n_s$ is the number of synchronized rods, and $N$ is the total number of rods. Indeed, we observe a sinusoidal oscillatory behavior of the rods in HS, as illustrated in Figs. \ref{fgr:DAF}(a) and \ref{fgr:DAF}(b). The amplitude ($n_s/N$) of the DAF is constant in time in the HS regime, with  $n_s/N \gtrsim 0.62$, and it corresponds to the fraction of particles oscillating synchronously with the magnetic field. Fig. \ref{freqdist}(d) shows that even in the HS regime, the rods exhibit rotational motion with a non-zero phase difference that increases with increasing aspect ratio due to the (implicit) friction with solvent (Eq. \ref{eqfri2}) and the presence of neighboring rods. Additionally, higher driving frequencies lead to higher phase differences for fixed $B_0$ and $l$ due to increased friction from increased rod rotational friction coefficient and rod angular velocity (Eq. \ref{eqfri2}), since $\omega \approx \omega_0$ in the HS regime. Therefore, the synchronization occurs out of phase with $\mathbf{B}$ \cite{SJager}.

In the Intermediate Synchronization (IS) regime, the DAF may present a sinusoidal time dependence with constant amplitude such that $n_s/N < 0.62$ [cf. Fig. \ref{fgr:DAF}(c)] or a damped oscillatory behavior with time, as shown in Fig. \ref{fgr:DAF}(c). In addition, the distribution $f(\omega)$ presents an asymmetric double-peaked behavior due to the in-plane oscillations of the rods at the same time they slowly rotate following the field, which is a consequence of the competition between the now comparable rod-rod and rod-field interactions influenced by the increased friction. Indeed, the rod-field interaction, which generates the rotation of the rods, is effectively weakened due to the higher rotational friction in the IS regime. The slow effective angular velocity of the rods results in a distribution of phase differences ($f(\varphi_0)$) between each rod and the magnetic field. Note that as $l$ increases the positive and negative peaks in  $f(\omega)$ become more symmetric and the distribution ($f(\varphi_0)$) becomes more homogeneous [cf. Fig. \ref{freqdist}(e)], as the rotation of rods with larger aspect ratio $l$ relative to the field is suppressed. 

In the Low Synchronization (LS) regime, 
the influence of the magnetic field is significantly weaker than the rod-rod interaction. This is due to the low intensity of the magnetic field and considerable rotation friction caused by the large $\omega_0$ and increased $l$. As a result, we observe typically clustered phases in this scenario. The angular velocity distribution is centered around $\omega = 0$, as illustrated in Fig. \ref{freqdist}(c), indicating that the rods rotate in both directions equally, i.e., the rods exhibit angular oscillations as a response to the magnetic field, becoming less pronounced as the aspect ratio increases. The dipole auto-correlation function in the LS regime presents a monotonic decay with time, because the rods are clustered and do not follow the field [Fig. \ref{fgr:DAF}(d)]. Hence, the distribution of the phase differences are equally likely in the LS regime [Fig. \ref{freqdist}(f)].

\subsection{Microstructure Characterization}
Some configurations, even though belonging to the same synchronization regime, have distinct features in their microstructure. For the spatial distribution properties, we consider the pair correlation function \cite{rapaport2004art}, defined in $2D$ as
\begin{equation}\label{gdr}
\mathbf{g(r)} = \frac{\Braket{\sum_{a}\sum_{b\neq a}^N\delta(r -R_{ab})}}{2N\pi r \rho},
\end{equation}
\noindent where $R_{ab}$ is the separation between the centers of rods $a$ and $b$ (see Fig. \ref{Fig1}).

We also study the appearance of orientational order, i.e., the polarization, by analyzing the magnetic order parameter induced by the magnetic field \cite{alvarez}
\begin{equation}\label{mag}
M = \Braket{\frac{1}{N}\sum_i^N \boldsymbol{\hat{\mu}}_i(t) \cdot \boldsymbol{\hat{B}}(t)}\textrm{,}
\end{equation}
where $\boldsymbol{\hat{\mu}}$ and $\boldsymbol{\hat{B}}$ are unit vectors defining the orientations of the dipole moment and magnetic field, respectively. $M = 1$ indicates that all rods rotate in phase and synchronously with the field, i.e., the system presents a paramagnetic order. On the other hand, $M < 0$ indicates a diamagnetic order.

We also quantify the extent of aggregation between dipolar rods through polymerization, which measures the amount of rods bonded to at least one other. We consider two rods bonded if the shortest separation between them is $\leq 1.4\sigma$. Such a geometrical criterion was defined previously \cite{jorgedomingos,JorgeSM}, and it is related to the distance between a pair of rods that minimizes their interaction energy. We define the polymerization as the ensemble average of the ratio between the number of clustered rods, $N_c$, and the total number of rods, $N$
\begin{equation}\label{polime}
\Phi = \Braket {\frac{N_{c}}{N}}\textrm{.}
\end{equation}

In the following subsections, we discuss the properties of the different microstructures observed in each synchronization regime. 

\subsubsection{Patterns in the HS Regime}

\begin{figure}[]
\centering
  \includegraphics[height=8.5cm]{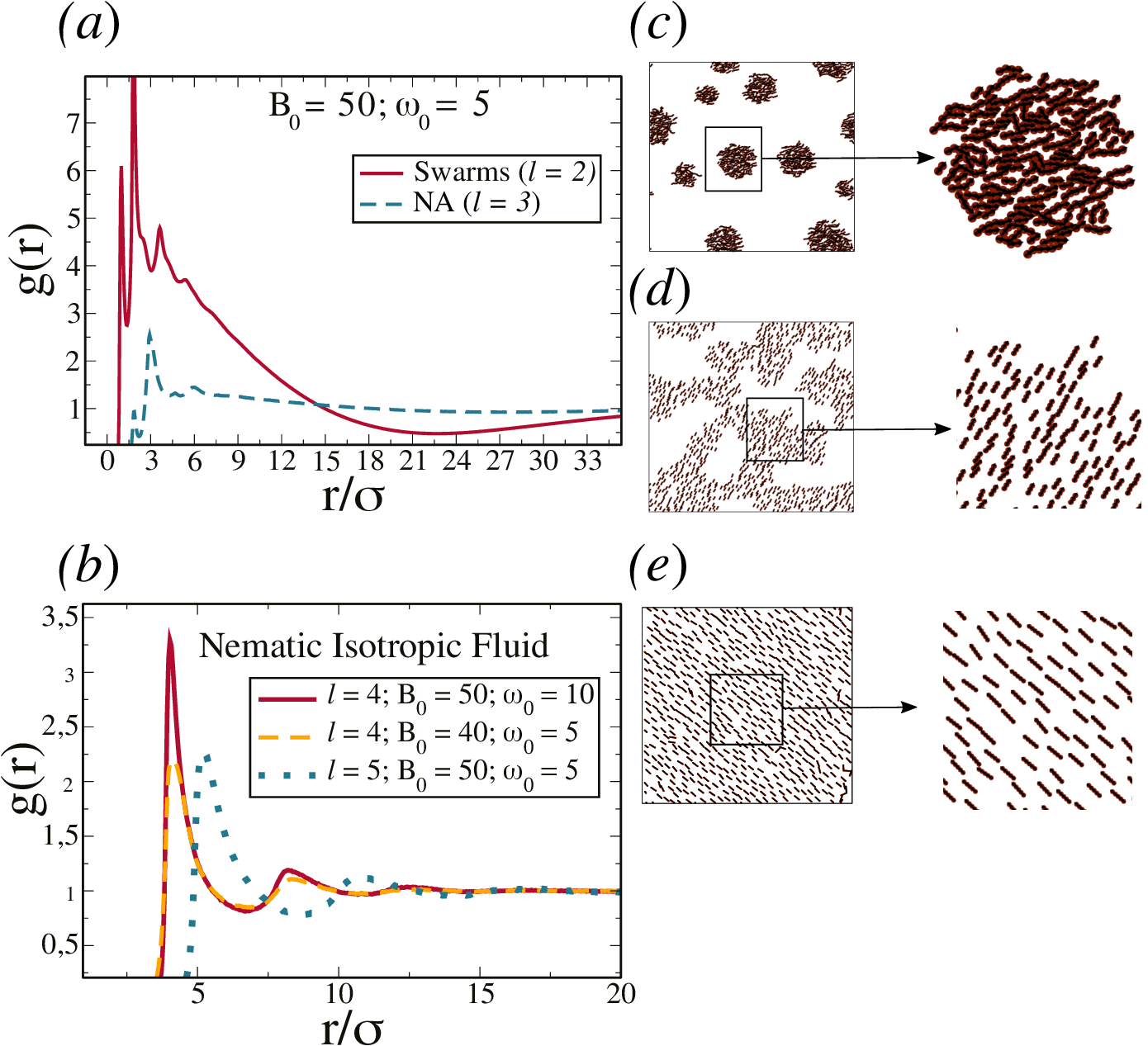}
  \caption{Pair correlation function of HS patterns: (a) Swarms and Nematic Aggregates; (b) Nematic Isotropic Fluid. Examples of the patterns are shown in: (c) Swarms; (d) Nematic Aggregates; and (e) Nematic Isotropic Fluid.} 
  \label{fgr:gdrHS}
\end{figure}

Since the majority of rods rotate synchronously with the field in the HS regime, we can understand the relationship between the observed synchronous states and the emergent clustering process by considering the time-averaged (over a period $\tau$) pair-interaction potential (Eq. \ref{eqdip}) of rods sync with the field. The result is an effective isotropic and purely attractive potential (Eq. \ref{inversedipo}) for $2D$ systems, 
\begin{equation}
u^D(\boldsymbol{r}_{ij}) = \frac{1}{\tau}\int_{0}^{\tau}u(\boldsymbol{r}_{ij},\boldsymbol{\mu}_i(t),\boldsymbol{\mu}_j(t))dt = - \frac{\mu^2}{2r_{ij}^3}\textrm{.}\label{inversedipo} 
\end{equation}
\begin{figure*}[]
\centering
  \hspace{-15.pt} \includegraphics[height=5.6cm]{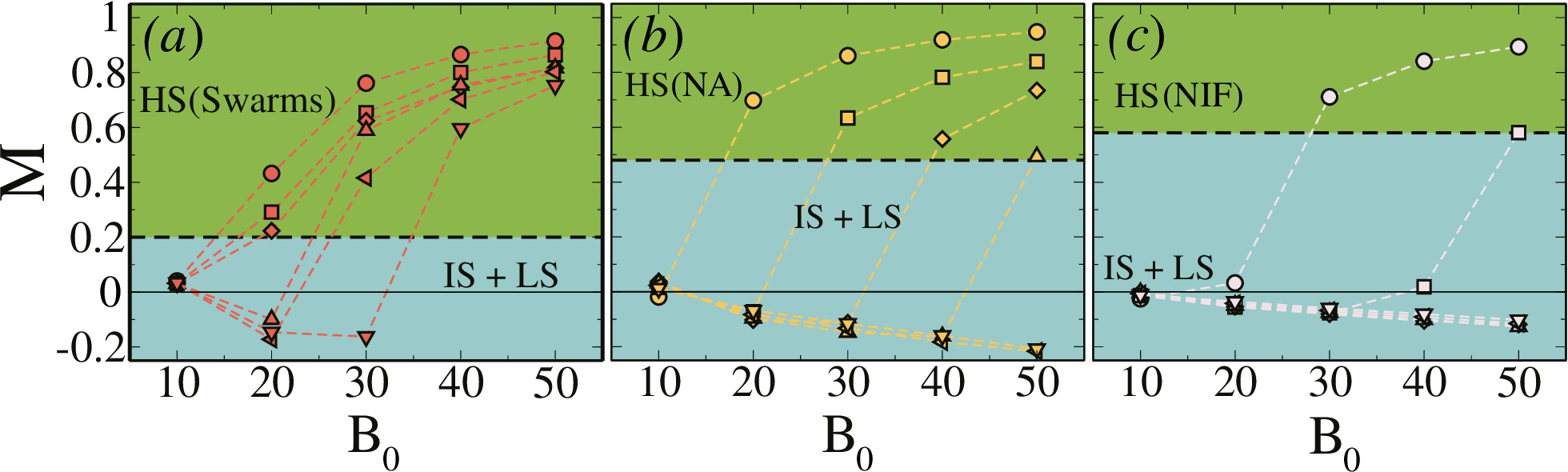}
  \caption{Magnetic order parameter as a function of the magnetic field $B_0$ for different aspect ratios of the rods: (a) $l = 2$; (b) $l = 3$; (c) $ l = 4$. Different symbols stand for different rotation frequency $\omega_0$, namely: $\protect\mycirc{white}$ for $\omega_0 = 5$; $\Box$ for $\omega_0 = 10$; $\Diamond$ for $\omega_0 = 15$; $\protect\mytrgle{white}$ for $\omega_0 = 20$; $\lhd$ for $\omega_0 = 25$; and $\triangledown$ for $\omega_0 = 30$. The HS regime is highlighted in green, while the non-HS regime is denoted by the blue region.}
  \label{fgr:MAG}
\end{figure*}

In the HS regime, most rods rotate in sync with the field, causing some attraction among them, which can, under certain conditions, lead to the formation of clustered structures. We observe three distinct patterns: Swarms, Nematic Aggregates (NA), and Nematic Isotropic Fluid(NIF). In Fig. \ref{fgr:gdrHS}, we show representative examples of the patterns observed in the HS regime for distinct aspect ratios, $B_0$ and $\omega_0$. 

The Swarms are observed only for $l =2$ in the HS regime. Their main structural feature is the presence of cohesive circular-like clusters (see Fig. \ref{fgr:gdrHS}(c)), which breaks the spatial isotropy of the system. Such behavior is seen in the decay of the pair correlation function, Fig. \ref{fgr:gdrHS}(a), marked by several unequal peaks at short distances, $r/\sigma\leq 6$, as a result of the typical separations inside the swarm structure,  and a decrease at larger distances, at which the minimum observed corresponds to the average separation between clusters ($r/\sigma \approx 21$).

As aforementioned, Eq. \ref{eqfri2} indicates that the rotational friction increases with increasing aspect ratio, and such a feature plays an important role in determining the resulting dynamic configurations. For example, we observe arrangements like Swarms for $l = 2$. In contrast, for $l = 3$, we find more dispersed configurations [Fig. \ref{fgr:gdrHS}(a)], but still with an anisotropic distribution of rods, namely, the Nematic Aggregates [Fig. \ref{fgr:gdrHS}(d)]. 
 For $l>3$, configurations with isotropic distribution of rods are observed, such as the Nematic Isotropic Fluid [Fig. \ref{fgr:gdrHS}(e)], a polarized fluid with a characteristic liquid behavior [Fig. \ref{fgr:gdrHS}(b)]. 

Concerning the magnetic ordering of the system, we also find that the observed microstructures have different polarizations in the HS regime, which can be seen through the magnetic order parameter, $M$, defined in Eq. \ref{mag} and shown in Fig. \ref{fgr:MAG}, where we present $M \times B_0$ phase diagrams for distinct values of driving frequency, and for systems with different aspect ratios ($l = 2, 3, 4$). In general, the synchronization of the rods with $\mathbf{B}(t)$ is more effective in low rotation, and we could expect, for a given $l$, that the lower $\omega_0$, the larger $M$ for any value of $B_0$. However, the synchronization is typically out of phase [see Fig. \ref{freqdist}(d)], resulting in some patterns in the HS regime exhibiting a relatively low value of $M$. For example, the polarization of Swarms for $l = 2$, $B_0 = 20$, and $\omega_0 = 15$ is $M \approx 0.2$  [cf. Fig. \ref{fgr:MAG}(a)], indicating that 
the synchronization driven by the field does not necessarily imply an intense polarization aligned with $\mathbf{B}(t)$. Moreover, several structures exhibit negative values of $M$ in the IS regime, due to in-plane oscillations of the rods while they slowly rotate following the field. The average phase difference between the rods and the field is $>\pi/2$ [cf. Fig. \ref{fgr:MAG}(e)], resulting in a negative value of $M$ in these cases. Patterns in the LS regime present $M \approx 0$. 

The magnetic order parameter defined in Eq. \ref{mag} describes the average component of the resultant magnetic moment of the system along $\mathbf{B}(t)$. The underlying tendency observed for $M$ as $l$ increases is the requirement of a stronger and slower field to induce more polarized patterns in the HS regime, as seen in Figs. \ref{fgr:MAG}(b) and \ref{fgr:MAG}(c) above the dashed lines; the approximate boundary between the dynamical regimes is higher for larger $l$, so that only slow fields ($\omega_0 = 5$) provide paramagnetic structures for $l=4$. Additionally, negative polarizations are observed for lower frequencies in the systems with larger $l$. Note that, for $l = 2$, negative polarization occurs only for $B_0 < 30$. While, as $l$ increases, negative polarization is obtained for any value of $B_0$. For given $B_0$ and $l$, the effective polarization of the system can be altered between positive and negative values through the rotating frequency of the magnetic field, suggesting the possibility of switching between two well-defined magnetic states.

\subsubsection{Patterns in the IS Regime}
 In the IS regime, we observe only a single pattern, the Isotropic Fluid. This structure is characterized by a liquid-like phase which clusters are absent, and a uniform particle distribution prevails throughout the system.
\begin{figure}[h!]
 \hspace{-15.pt} \includegraphics[height=4.cm]{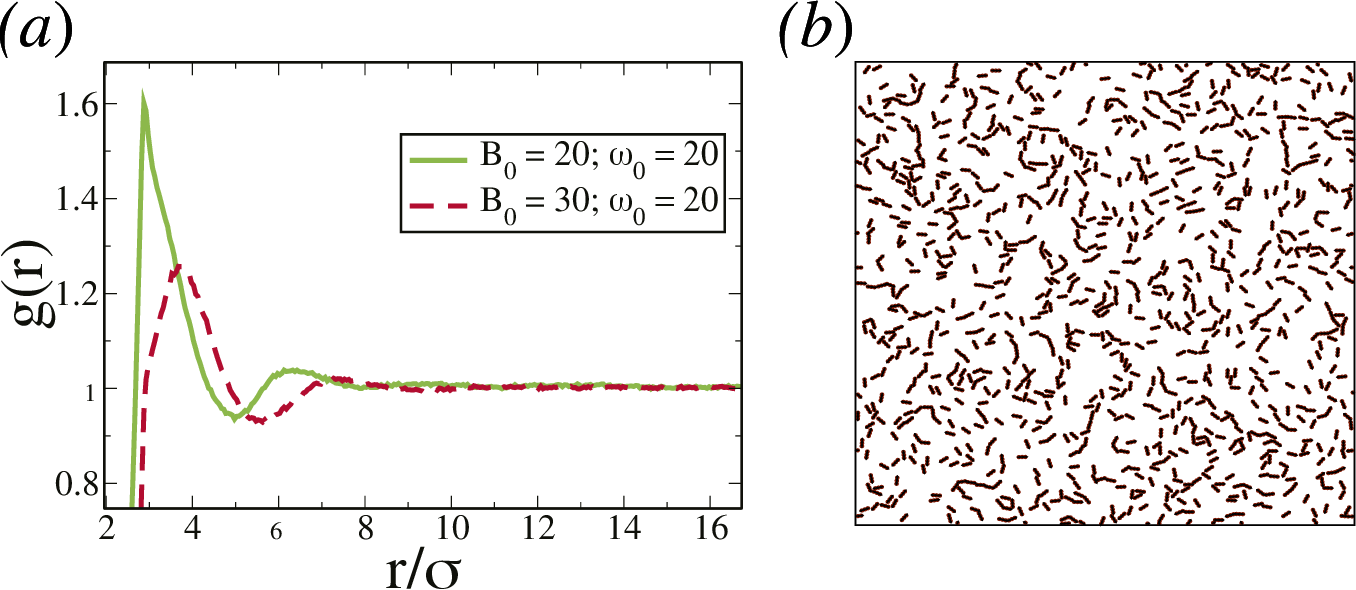}
\caption{(a) The pair correlation function for $l = 3$ in the IS regime. (b) Snapshot from our computer simulations obtained for $B_0 = 20$ and $\omega_0 = 20 $.}
  \label{fgr:Iscorra}
\end{figure}
The Isotropic Fluid phase shares structural characteristics with the Nematic Isotropic Fluid, exhibiting low polymerization and liquid-like behavior, as seen in the pair distribution function, $g(r)$, displayed in Figs. \ref{fgr:gdrHS}(b) and \ref{fgr:Iscorra}(a). However, the Isotropic Fluid lacks magnetic order due to the high friction observed in this phase, thus preventing the proper alignment of rods along the magnetic field direction and resulting in a large phase difference.

\subsubsection{Patterns in the LS Regime}\label{subsecLS}
 
\begin{figure}[h!]
 \hspace{-50.pt} \includegraphics[height=7.7cm]{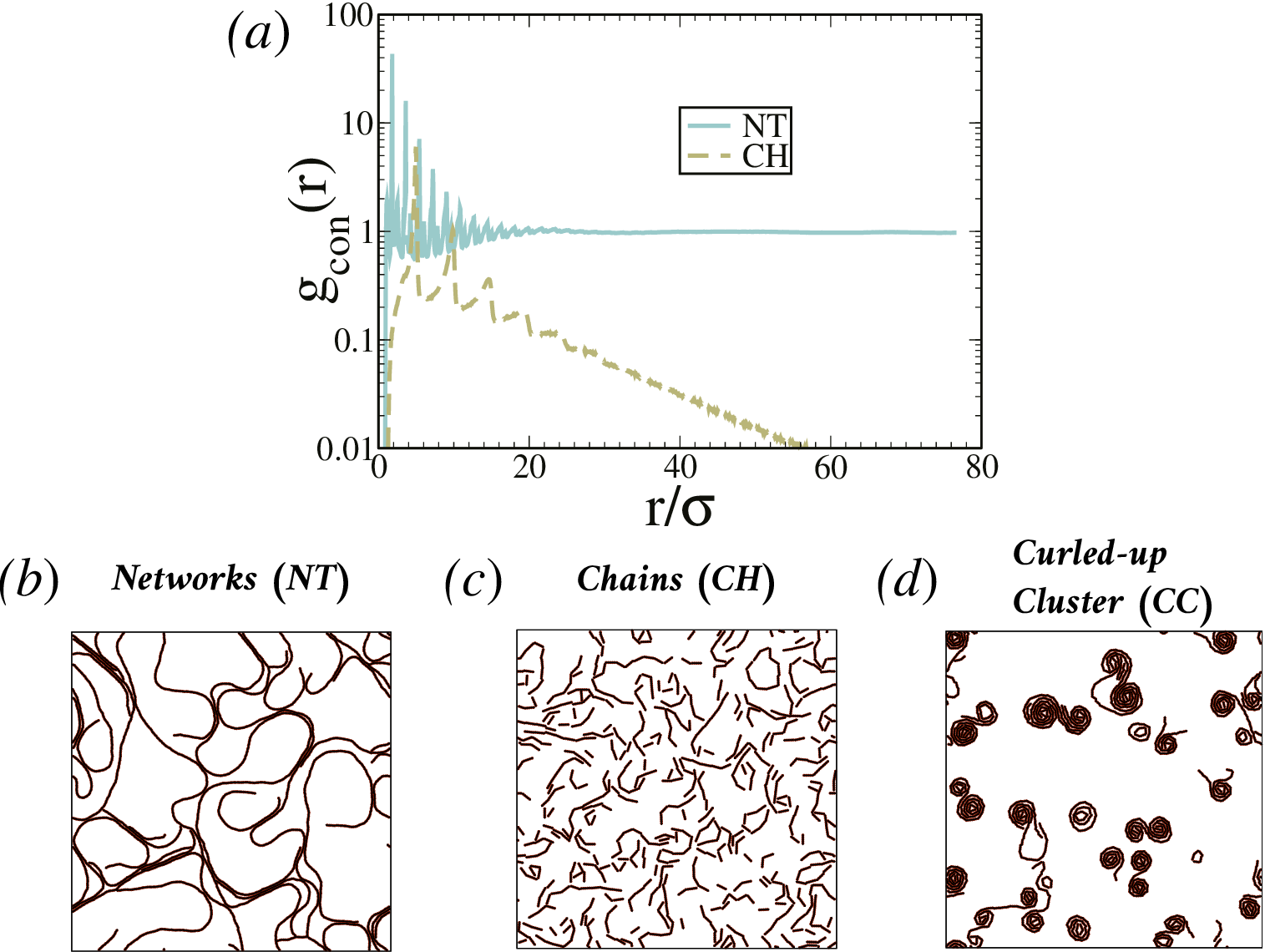}
\caption{Low Synchronization (LS) regime: (a) pair connectedness correlation function for Networks (NT): $l = 2$; $B_0 = 10$; $\omega_0 = 20$, and Chains (CH): $l = 4$; $B_0 = 10$; $\omega_0 = 20$. The $g_{con}(r)$-axis is in the log scale, while $r/\sigma$-axis is in the normal scale.  (b)-(d) Illustration of different patterns observed in the LS regime. The Curled-up Cluster (CC) for $l = 2$; $B_0 = 10$; $\omega_0 = 5$ is shown in (d).}
  \label{fgr:ls}
\end{figure}

In the Low Synchronization (LS) regime, the colloidal particles self-organize into a network of wormlike chains, a common characteristic of dilute, strongly interacting dipolar systems \cite{jordanovic}. However, our simulations reveal three distinct microstructures (see Fig. \ref{fgr:ls}), each exhibiting different connectivity properties depending on the aspect ratio of the rods. These microstructures are Networks (NT), Chains (CH), and Curled-up Clusters (CC).

\begin{figure}[h!]
 \hspace{0.pt} \includegraphics[height=5.7cm]{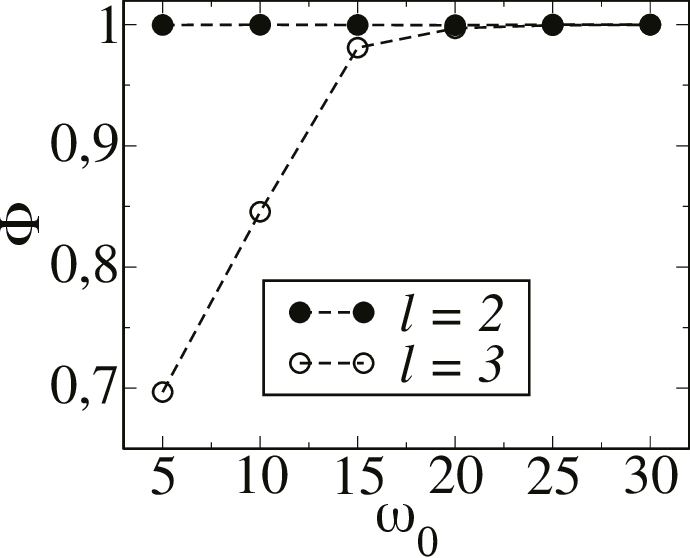}
\caption{Degree of polymerization as a function of frequency $\omega_0$ for $B_0 = 10$ and different aspect ratios.}
  \label{fgr:polimB10}
\end{figure}
NT are characterized by head-to-tail arrangements of rods forming spanning clusters. In cases where those head-to-tail aggregates only result in small chains, we have CH microstructures. On the other hand, CC microstructures consist of chain-like clusters that present larger chains that CH, but are not large enough to exhibit percolation. As a result, the clusters curl into distinctive configurations under the influence of the magnetic field. Therefore, the length of resulting chains is an explicit distinction among the NT, CH, and CC microstructures. 

To check the percolation, we employ the pair connectedness correlation function, similar to Eq. \ref{gdr}, but applied to sequential bonds \cite{sciortino}. A vanishing $g_{con}(r)$ as $r/\sigma \rightarrow \infty$ indicates a lack of percolation, as illustrated in Fig. \ref{fgr:ls}(a), where we compare the $g_{con}(r)$ functions for microstructures NT and CH.

The aspect ratio of the rods, intensity, and rotation frequency of the external field are relevant parameters to set the dynamics and structure of the systems, as in the LS regime. For example, we find rigid NT configurations for $l = 2$, $B_0 = 10$, and $\omega_0 \geq 10$. At the same time, for $\omega_0 = 5$, a more soft CC microstructure is observed, showing the potential to control the structure and dynamics of the system with an external rotating field. For $B_0 = 10$ and $\omega_0 = 10$, a system with $l = 3$ self-organizes into a CH configuration, which is similar to the one with $l = 2$ (NT), but not rigid and not percolated, while for $l \geq 4$ the system self-organizes into an Isotropic Fluid.    
The formation of Curled-up Clusters differs according to the aspect ratio. For $l = 2$, the CC is observed only for $\omega_0 = 5$ due to the stronger correlation with the magnetic field responsible for curling the chains. On the other hand, still for $l = 2$, the formation of spanning clusters in the LS regime, observed for $\omega_0\geq 10 $, results in rigid clusters, which are characteristic of low-density gels sustained through attractive interactions due to rigidity percolation \cite{zhang}. However, for larger aspect ratios ($l=3$ and $l=4$), the curling process happens for larger $\omega_0$, in contrast to the $l = 2$ case, since there are no rigid clusters due to the absence of percolation. In such cases, the polymerization for $B_0 = 10$ increases as $\omega_0$ increases (see Fig. \ref{fgr:polimB10}), enabling the clusters to get large enough to be curled up. The difference in rigidity of patterns in LS can also be analyzed by the individual rotational dynamics revealed by the decay rate of DAF in Fig. \ref{fgr:DAF}(d).


\subsection{Phase Diagram}
\begin{figure*}[t!]
\centering
  \includegraphics[height=9.2cm]{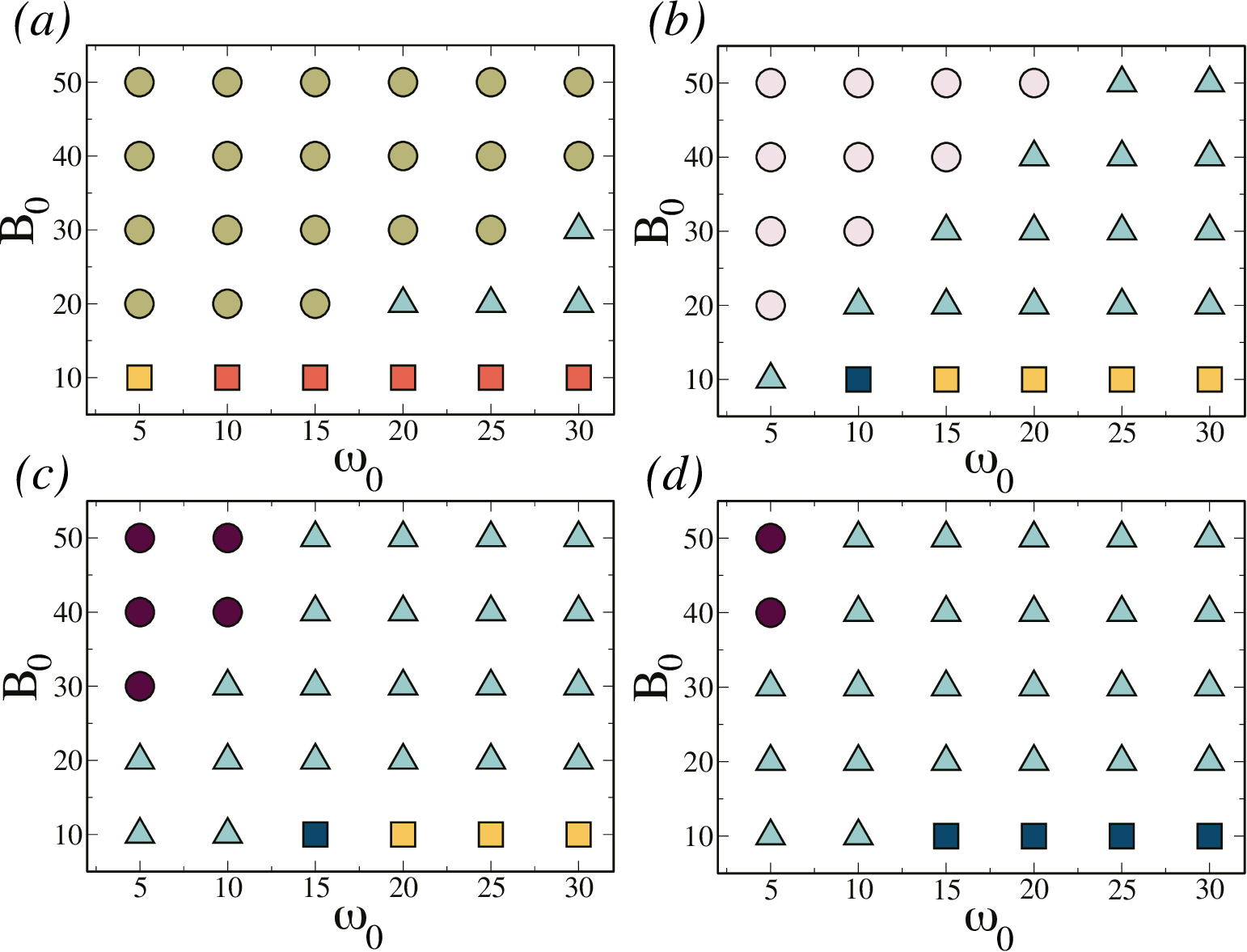}
  \caption{Steady state phase diagrams presenting the self-organized structures as a function of the intensity, $B_0$, and the rotation frequency, $\omega_0$, of the magnetic field for different values of the aspect ratio: (a) $l = 2$; (b) $l = 3$; (c) $l = 4$; (d) $l = 5$. Symbols represent different regimes of synchronization: \protect\mycircle{white} High Synchronization, \protect\mytriangle{white} Intermediate Synchronization, and \protect\mysquare{white} Low Synchronization. Different colors stand for different microstructures in each synchronization regime, i.e., \protect\mycircle{mygreen} Swarms, \protect\mycircle{mypink} Nematic Aggregates, \protect\mycircle{mywhitepeach} Nematic Isotropic Fluid, \protect\mytriangle{myblue} Isotropic Fluid, \protect\mysquare{myred} Networks, \protect\mysquare{mymaroon} Chains, and \protect\mysquare{myyellow} Curled-up Cluster.}
  \label{fgr:phasediagram}
\end{figure*}
We summarize the observed configurations in phase diagrams presented in Fig. \ref{fgr:phasediagram} for different $l$. Different symbols in the phase diagram illustrate the three distinct synchronization regimes: \mycircle{white} for the High Synchronization, \mytriangle{white} for the Intermediate Synchronization regime, and \mysquare{white} for the Low Synchronization regime. Additionally, we use different colors to indicate specific patterns within each synchronization regime. For the HS regime, we observe three patterns: \protect\mycircle{mygreen} Swarms, \protect\mycircle{mypink} Nematic Aggregates, and \protect\mycircle{mywhitepeach} Nematic Isotropic Fluid. In the IS regime, we observe only the Isotropic Fluid, represented by the symbol \protect\mytriangle{myblue}. In contrast, the LS regime is characterized by: \protect\mysquare{myred} Networks, \protect\mysquare{mymaroon} Chains, and \protect\mysquare{myyellow} Curled-up Clusters. 

One crucial factor influencing the synchronization process with the magnetic field is the aspect ratio of the rods. The effect of distinct $l$ on the phase diagrams is evident, as we observe a drastic reduction in the High Synchronization region (circles) as $l$ increases. Conversely, there is a corresponding enlargement of the Intermediate Synchronization region (triangles) as we compare phase diagrams for different aspect ratios. The general trend in these plots reveals a consistent pattern: synchronization tends to decrease as the frequency increases, whereas larger magnetic fields lead to enhanced synchronization. Both factors directly influence the rotational motion by affecting the total torque in each rod. The broad range of the IS regime can be attributed to an intricate interplay between the rotational motion induced by the magnetic field and the effect of the increase in total friction. In that specific region of the phase diagram, as mentioned earlier, we observe that the resultant effect is that the magnetic field effectively hinders clustering. Furthermore, the expansion of the IS regime with increasing aspect ratio is also attributed to the resulting increase of rotational friction, as outlined in Eq. \ref{eqfri2}, acting as a suppressor of synchronization. This friction plays a crucial role in modulating the extent of synchronization. Consequently, most phase diagrams, for $l > 2$, are in the Isotropic Fluid phase, characterized by poor alignment, low magnetic order ($M$), and the absence of spatial order.

\section{\label{summary}Summary and Conclusions}

We investigated the pattern formation of a two-dimensional magnetic peapod rod system subjected to rotating magnetic fields through molecular dynamic simulations. Each rod was composed of soft beads, each having a central point-like dipole whose aspect ratio is numerically equal to the number of beads. We studied how the system behaves based on the strength of the magnetic field, its rotation frequency, and the aspect ratio of the rods. A key aspect of our study was the synchronization process, which emerged due to the friction and the competition between rod-rod and rod-field interactions. We showed a significant synchronization suppression with increasing aspect ratio of the rods and driving frequency, which gave rise to several emergent patterns in three specific synchronization regimes. Our choice of magnetic field values aligns with typical field-driven experiments. Although consistent with other numerical studies \citep{Jaeger,SJager,JorgeSM}, the frequency values were notably higher than those in real-world systems. This divergence could be attributed to the many-particle character of hydrodynamic interactions induced by the solvent, which affects the system's dynamics \cite{SJager}.

The system exhibited three distinct synchronization regimes: High, Intermediate, and Low. In the High Synchronization, we observed three distinct patterns: Swarms, Nematic Aggregates, and Nematic Isotropic Fluid, which depended on the rod aspect ratio ($l$). Swarms, for instance, appeared as an anisotropic pattern, observed for $l = 2$, since the increased friction disfavors the formation of such patterns for higher $l$. Some spatial isotropy was observed for rods with a larger aspect ratio, followed by the emergence of a nematic phase, giving rise to the Nematic Aggregates. Further spatial isotropy was achieved with an increasing aspect ratio, and a liquid-like polarized pattern was observed in the Nematic Isotropic Fluid. Each of these patterns, and their associated synchronization regimes, provide unique insights into the system's behavior under different conditions, contributing to our understanding of field-driven systems. 
 
Notably, the magnetic order was not a guaranteed outcome in the HS regime, as seen by the lowest value of the magnetic order parameter ($M \approx 0.2$) observed for Swarms. The analysis of the distribution of the phase differences revealed that, although the distribution of angular velocities is centered at the driving frequency $\omega = \omega_0$, the synchronization occurred out of phase because of friction.

In the Intermediate Synchronization regime, we observed a dispersed, non-polarized, and spatially isotropic pattern, the Isotropic Fluid. In this regime, the increased friction resulting from the increased rod aspect ratio and driving frequency led to a relative equivalence of the interactions caused by the weakening of the rod-field interaction compared to the rod-rod interactions. In this scenario, the field-driven rotation of the rods was observed to be very slow followed by in-plane oscillations. In some cases, the resulting average phase difference between rods and the external field was reflected in the magnetic order parameter $M$, where we observed a negative order concerning the direction of the magnetic field. More interestingly, such effective positive and negative magnetic order could be tuned according to the frequency of the external field, indicating the exciting possibility of controlling such field-driven systems, a prospect that holds great promise for future applications.  

In the Low Synchronization regime, characterized by low magnetic field values and high rotation frequencies, the interaction among magnetic rods was dominant over the rod-field interaction. The response to the magnetic field was limited to local oscillations through the bonds formed by the rods, as revealed by the distribution of angular velocities  Thus, clustered phases emerged with chain-like head-to-tail arrangements, displaying three distinct patterns with varying connectivity properties based on the length of the resulting chains: Networks, Chains, and Curled-up Clusters. Notably, due to the concept of rigidity percolation, which refers to the point at which a rigid cluster spans the entire system, the appearance of rigid clusters was limited to the $l=2$ case. This phenomenon has significant implications for the system's structural properties, as it determines the system's ability to maintain its shape under external forces. Furthermore, for the $l = 2$ case, the Curled-up Cluster was only observed for $B_0=10$ and $\omega_0=5$. Conversely, for larger aspect ratios, the curling process exhibited distinct behavior. Curled-up Clusters were not observed for $l = 5$, but their presence became apparent for higher driving frequencies, as the degree of polymerization increased with $\omega_0$, for $l=3$ and $l=4$.

As we presented the comprehensive $B_0 \times \omega_0$ phase diagrams, outlining the conditions for specific self-organized structures emerge, we also pave the way for further exploration. Our research has provided valuable insights into the synchronization-driven self-organization of magnetic rod systems, but also raises intriguing questions. Unexplored aspects related to factors such as hydrodynamics, temperature, density, the system's dimensionality warrant further investigation, promising exciting pathways for future research.


%



\begin{acknowledgments}
This work was supported by the Brazilian agencies FUNCAP, FAPESPA, CAPES and CNPq.
\end{acknowledgments}


\end{document}